\documentclass[referee]{aa}

\usepackage{graphicx}
\usepackage{txfonts}
\usepackage{natbib}
\begin{document}

\title{Magnetic topologies and two-class coronal mass ejections:
a numerical magnetohydrodynamic study}

\titlerunning{Magnetic topologies \& two-class CMEs: a numerical MHD study}
\authorrunning{W. Liu et al.}

\author{W. Liu \inst{1,2} \and
 X. P. Zhao \inst{1} \and
 S. T. Wu \inst{3} \and
 P. H. Scherrer \inst{1} }

\offprints{W. Liu, \email{weiliu@quake.stanford.edu} }

\institute{W. W. Hansen Experimental Physics Laboratory, Stanford University,
  Stanford, CA 94305, USA   
  \and
  Department of Physics, Stanford University, Stanford, CA 94305, USA
  \and
  Center for Space Plasma and Aeronomic Research, 
  Department of Mechanical and Aerospace Engineering, 
  The University of Alabama in Huntsville, Huntsville, AL 35899, USA \\
  }

\date{Received 14 October 2004 / accepted 20 December 2004}

\abstract{
White-light observations of the solar corona show that there are 
two characteristic types of Coronal Mass Ejections (CMEs) in terms of
speed-height profiles: so-called
fast CMEs that attain high speeds low in the corona and slow
CMEs that gradually accelerate from low initial speeds.  \citet{low02}	
have recently proposed that fast and slow CMEs result from initial 
states with magnetic configurations characterized by normal prominences (NPs) and 
inverse prominences (IPs), respectively.  To test their theory,
we employed a two-dimensional, time-dependent, resistive 
magnetohydrodynamic code to simulate the expulsion of CMEs in these
two different prominence environments.  Our numerical simulations demonstrate
that (i) a CME-like expulsion is more readily produced in an NP
than in an IP environment, and, (ii) a CME originating
from an NP environment tends to have a higher speed 
early in the event than one originating from an IP
environment.  Magnetic reconnection plays distinct roles in the 
two different field topologies of these two environments to produce
their characteristic CME speed-height profiles.  Our numerical simulations
support the proposal of \citet{low02}  
although the reconnection development
for the NP associated CME is different from the one
sketched in their theory.  Observational implications of 
our simulations are discussed.

   }

\maketitle

\section{Introduction}

The accumulated observations of Coronal Mass Ejections (CMEs) 
obtained over more than 3 solar
cycles by the coronagraphs onboard the {\it Skylab}, {\it Solar Maximum Mission},
and {\it Solar and Heliospheric Observatory} ({\it SOHO}) have shown 
two characteristic types of CME speed-height profiles:
(I) so-called fast CMEs attaining high speeds 
above the CME median speed (400 km s$^{-1}$) low in the corona 
with little or even negative subsequent accelerations, and, 
(II) slow CMEs gradually accelerating from low initial speeds 
($\leq 400$ km s$^{-1}$)
\citep{gos76, mac83, dry96, she99, cyr99, cyr00, moo02}. 
It has also been recognized that 
Type I CMEs tend to originate from active regions and are frequently accompanied by 
flares, while Type II events usually originate away from active regions 
and are accompanied by eruptive prominences.
Outstanding examples of these two characteristic types can be 
found in a given sample of CMEs, but it should be pointed out that 
there is a continuous spectrum of speed-height profiles between these
two types as the opposite extremes. 
Another way to distinguish these two types of events was given by
\citet{and01}. They showed that CMEs observed by the LASCO 
coronagraphs onboard {\it SOHO} may fall into two categories: 
Type C characterized with a constant
speed and Type A characterized with a significant acceleration, 
which  respectively correspond to Types I and II.

The underlying physics responsible for the dual 
character of CME speed-height profiles remains among the 
outstanding questions in CME research. 
Recently, Low \& Zhang (2002, hereafter referred to as LZ02)
suggested a theory to explain this phenomenon
in terms of the different hydromagnetic environments in which 
CMEs occur. CMEs are correlated at about the 75\% level with 
prominence eruptions \citep{mun79, web87, cyr99}.
There are two magnetic types of quiescent prominences referred
to as the Normal and Inverse Prominences \citep{tan95}, 
hereafter called NPs and IPs respectively.
Low \& Zhang pointed out that these two types of prominences
represent different magnetic field topologies which have
distinct consequences for the interplay between magnetic reconnection
and CME expulsion dynamics. They gave qualitative sketches
of this hydromagnetic interplay which suggest that fast and slow CMEs 
are naturally associated with NPs and IPs, respectively.

In the case of an NP, the surrounding 
field has a topology such that the CME is expelled with 
a current sheet to be dissipated by 
magnetic reconnection ahead of the erupting prominence.
This reconnection is a break-out effect similar to the
one originally proposed by \citet{ant99}
for a multipolar magnetic field. In the LZ02 proposal, the global field
is bipolar with a magnetic flux rope \citep{che89,che97}. The rise of the 
flux rope and prominence drives break-out reconnection 
ahead. Reconnection produces a slingshot effect which, 
in turn, drives the CME and prominence. This runaway situation
naturally produces a fast CME with an impulsive acceleration and flare
heating early in the event. It is worth mentioning that such a timing correlation
between a fast CME and flare does not necessarily mean that the flare
is driving the CME.

In contrast, an IP is associated with a magnetic-flux
rope topology which produces a current sheet trailing behind the
CME and erupting prominence.  This current sheet is not driven directly
by the CME but forms passively from the left-behind magnetic field.
Magnetic reconnection is then not compelled to occur early and 
there is no impulsive CME acceleration or flare heating
by reconnection early in the event.  A CME produced
in this hydromagnetic environment may be expected to be gradually
accelerated from a low initial speed.

Studies of flare morphology and timing associated with fast and slow CMEs
\citep{zha02, zha03}
have shown that observations of this kind are consistent with the 
proposed theory of LZ02.  Ultimately, observations of more incisive
kind will be needed to verify or reject the LZ02 theory.
On the other hand, the proposed theory of LZ02 is based on intuitive
sketches of the relevant hydromagnetic processes.  It is therefore
important to perform numerical magnetohydrodynamic (MHD) 
simulations to directly investigate if the two distinct CME 
acceleration processes of LZ02 can be produced.

Individual simulation works have been reported in the literature on 
the acceleration of CMEs.  
\citet{wu95} and \citet{wu97},
respectively using a 2-D and 2.5-D ideal MHD model, studied CME expulsions involving 
a magnetic flux rope that can be interpreted to be those representing an IP. 
In their models, there was no magnetic reconnection taking place. 
A similar study, but involving reconnection in a configuration resembling an NP,
was performed by 
\citet{guo96} using a 2-D resistive code.  
However, the topological differences between the two types of configuration
had not been recognized in connection with the two-class CMEs by these
authors, nor by others until the recent		
work by LZ02.  On the other hand, different numerical models adopted in these 
works \citep{wu95, wu97, guo96} make it difficult to directly compare the CME 
acceleration between the two configurations from their results.
Therefore a systematic, comparative simulation study of 
CME expulsions involving magnetic reconnection in both IP and 
NP configurations will be instructive in checking on
the ideas of LZ02 and in an exploration of the interplay between
magnetic reconnection and CME expulsion dynamics.  Such a study is
reported in this paper, and, as we shall see, new effects 
not considered by LZ02, with regard to CMEs in an NP	
environment, are among the results we will report.
\S\ 2 describes the numerical model.
Simulation results are presented in \S\ 3,
followed by discussions on operating forces, magnetic topologies,
and magnetic reconnection in \S\ 4. Finally, we conclude this paper in \S\ 5 
with remarks relevant to the LZ02 theory and observational implications.


\section{Description of the Simulation Model}

The numerical model used for this study, based on those of \citet{wu95} 
and \citet{guo96}, is composed of a set of two-dimensional, time-dependent, 
resistive, single-fluid MHD equations in the spherical coordinates 
($r$, $\theta$, $\phi$), 
under an axisymmetry assumption ($ {\partial}/{\partial \phi}=0$).
The governing equations, including the conservation laws of mass, 
momentum, and energy and the magnetic induction equation, are identical to 
those in \citet{guo96} which the interested reader is referred to for the 
mathematical formulation.  We summarize the model as follows.

(i) The computational domain is defined as $1 R_s \leq r \leq 7.14 R_s$
and $-1.5^{\circ} \leq \theta \leq 91.5^{\circ}$, 
where $R_s $ is the solar radius and $\theta=0^{\circ}$ ($\theta=90^{\circ}$) 
the north pole (equator). We used an $81 (r) \times 63 (\theta)$ grid,
uniform in the meridional direction ($\Delta\theta=1.5^{\circ}$) 
and non-uniform in the r-direction
($\Delta r_i \equiv r_i - r_{i-1} = 0.95 [\Delta\theta] r_{i-1}$)
for a better resolution near the solar surface, with $\Delta r_i$ ranging
from $0.025 R_s$ at the coronal base to $0.178 R_s$ at the outer boundary. 
The grid was staggered to prohibit sawtooth oscillations.

(ii) The MHD equations were solved by the combined difference technique 
\citep{guo91, wu95}. The essence of this technique 
is to employ different numerical schemes to treat different equations 
according to their physical nature. Namely, 
the second-type upwind scheme is used for the continuity and energy
equation, the Lax scheme for the momentum equation,
and the Lax-Wendroff scheme for the magnetic induction equation.

(iii) The boundary conditions are: 
(a) a symmetric boundary at the equator and pole, 
(b) linear extrapolation at the outer boundary,
and (c) the method of projected normal characteristics \citep{wu87}
at the inner boundary.  To guarantee the solenoidal condition, 
viz., $\nabla \cdot {\bf B}=0$, the reiterative divergence-cleaning 
method \citep{ram83} was applied.

(iv) One of the major modifications to the previous model was that we
prescribed a uniform magnetic resistivity ($\eta =8.75 \times 10^3 \Omega$ m)
throughout the whole computational domain, 
rather than the anomalous resistivity used by \citet{guo96} 
which favors reconnection only in regions with large electric current density.
By doing this, various regions were treated equally, in view of the large 
difference in the current density distribution between the IP and NP 
configurations as we shall see in \S\ 4.  This modification is essential for 
a direct comparison between the two configurations.

We took three steps to simulate the expulsion of a CME in the corona, 
following \citet{wu95} and \citet{guo96}.
First we constructed an initial state of the corona with a quasi-equilibrium
helmet streamer by using the relaxation method \citep{ste82}.  
The resulting magnetic field, plasma flow velocity, and electric current 
density are shown in Figure 1 
(see Fig. 1b in Wu et al. 1995 for the corresponding plasma density distribution). 	
The characteristic parameters are listed in Table 1.

In the next step, we emerged a flux rope with various energy contents and two 
types of magnetic configuration	
from below the photosphere into the corona.  Our 2-D model approximates the
central cross-section of the 3-D flux rope which could be anchored at 
two ends on the photosphere in a realistic geometry.
The gas pressure and magnetic field of the flux rope 
(see Equations (4) and (5) in Guo et al. 1996), in an equilibrium state, 
were analytically specified in local cylindrical coordinates $(r', \theta',z')$,
with Equation (5) being modified:
\begin{equation}
{\bf B}(r')=\pm \mu_0 j_{0} ({\frac {1}{2}}a r'-{\frac {1}{3}} r'^2) {\bf
e}_{\theta '},
\end{equation}
where $a$ is the radius of the flux rope.
The ``$\pm$" in this equation represents different senses of circulation 
(or polarity) of the flux rope's magnetic field: the poloidal field
in the {\it upper} part of the rope with a ``+" (``$-$") sign is of the 
same (opposite) circulation with respect to the external coronal field. 
Considering that the prominence material is presumably contained in the 
{\it lower} part of the flux rope, the ``+" (``$-$") therefore corresponds 
to an IP (NP) topology \citep{tan95}. Equivalently, the azimuthal current in the
emerging flux rope flows in the same direction for the ``+" case as that 
in the vertical current sheet of the helmet streamer (Fig. 1b).
To implement the emergence process, we initially placed the flux rope
below the photosphere with its center 
at $r = R_s - a$. We then slowly displaced the rope upward at a 
constant speed $v_{em}$ ($\ll v_A$, see Table 2) by accordingly changing 
the physical variables at the inner boundary. 
It took 4 hours for the flux rope to entirely emerge into the corona
and then the inner boundary conditions were set back to their original form.
The reader is referred to \citet{wu95} for a detailed 
mathematical treatment of this flux rope emergence process.
The last step was carried out by simply letting the system evolve by itself until $t=40$ hrs.
We present the simulation results in the next section.


\section{Simulation Results}

A total of 8 simulation cases, grouped into 4 pairs, were performed with 
various sizes of the emerging flux rope.
The two cases of each pair have identical physical conditions except
opposite magnetic polarities of the emerging flux rope, 
respectively corresponding to the NP and IP topology.
The energy contents (i.e., the combined magnetic and thermal energy)
inside the flux rope (assuming the third-dimensional thickness 
$\Delta z' = 0.1 R_s$ in the local cylindrical coordinates) 
are $1.76 \times 10^{31}, 3.95\times 10^{31}, 
7.02\times 10^{31}$, and $1.10\times 10^{32}$ ergs, respectively for 
the 4 pairs of cases (cf. $\sim 10^{32}$ ergs needed to expel a moderately 
large CME, see Hundhausen 1999; Forbes 2000). 
We tabulate the key parameters of these cases 
in Table 2 and describe detailed results as follows.

Figure 2 shows the height-time and speed-time profiles
of the ``erupting flux rope"\footnote
{For the NP configuration, the ``erupting flux rope" refers to the 
new flux rope formed by reconnection; for the IP configuration, 
it refers to the originally emerging flux rope. 
This will be further explained in the following text.}
center (defined as the O-type neutral point)
for the studied cases.  The solid (dotted) lines represent the cases of the
NP (IP) environment. 
As we can see, the differences between the two types of environments are evident.
In Cases 1a and 2a, the emerging flux rope has already destabilized 
the helmet streamer and launched a CME by the end of the simulation
in the NP configuration.
In contrast, the streamer and the flux rope are still in equilibrium 
in the IP configuration in Cases 1b and 2b.
As expected from LZ02's theory, these profiles reveal distinct 
characteristics between the two types of topologies: 
(i) an NP environment seems to be more in favor of producing CMEs 
than an IP environment (also see Zhang \& Low 2003);   
(ii) a CME produced in an NP
configuration tends to have a higher speed in its early life than a
CME (if any) originating from an IP environment under otherwise
identical conditions. We take note of that 
the average CME eruption speed in Case 4a is about 2.3 times that 
in Case 4b (see Table 2), which agrees well with the typical 
fast-to-slow speed ratios of two-class CMEs 
derived from the {\it Skylab} and recent {\it SOHO}/LASCO coronagraphs 
(e.g. 775 km s$^{-1}$ versus 330 km s$^{-1}$ in Gosling et al. 1976,  
and 955 km s$^{-1}$ versus 411 km s$^{-1}$ in St. Cyr et al. 1999).	
The other cases in this study show much higher ratios.

Let us now focus our attention on two eruptive cases --- 4a 
(NP configuration) and 4b (IP configuration). 
Figure 3 shows the magnetic field and the plasma velocity
for these two cases. The corresponding azimuthal electric current density, 
$J_{\phi}$, and plasma density enhancement, 
$[\rho(r, \theta, t)-\rho(r, \theta, 0)] / \rho(r, \theta, 0)$,
are respectively displayed in Figures 4 and 5. 

\subsection{Eruptive Case --- Normal Prominence Configuration}

In Case 4a, we notice that upon the flux rope emergence, 
a curved current sheet is developed (Fig. 4a) between the leading edge of 
the emerging rope and the helmet dome (i.e., the  closed field region of 
the helmet streamer) whose bipolar magnetic field 
is tied on the photosphere and in a direction opposite to 
the field in the upper part of the flux rope. Due to the finite resistivity, 
magnetic reconnection occurs in the current sheet and rapidly forms a {\it new} 
flux rope between the {\it old} (emerging) flux rope and the helmet dome (Fig. 3a).
Once this new rope is formed, the current sheet splits into two halves (Fig. 4b)
and two X-type neutral points appear on the flanks of the old rope (Fig. 3b).
The magnetic flux in the new rope then grows 
as reconnection proceeds until all the closed-field flux in the 
helmet dome is converted to the new rope's flux.
In the meantime, the new flux rope expands 
and runs upward very fast while the old one follows and loses an equal 
amount of magnetic flux as the helmet-dome field above (Fig. 3b and 3c). 
Note that the old flux rope must contain sufficient
poloidal flux to annihilate the closed field of the helmet dome ahead
in order to set free the flux rope itself (also see LZ02). We identify 
the new flux rope as the main body of a CME in this numerical experiment. 
$v_f$ and $\bar{v}$ in Table 2 respectively refer to the final and average speed of 
the new flux rope's center for the cases of the NP configuration. 
At $t = 25$ hrs, the erupting material has escaped from the computational domain
with an average speed of $\bar{v}=161.0$ km s$^{-1}$ 
and the system reaches a quasi-equilibrium state (Figs. 3d and 4d)
similar to the initial state (Fig. 1).
We further note that, during the eruption phase (i.e., $t < 25$ hrs), 
the maximal current is on the flanks of the old
flux rope where the current sheets are located (e.g., Figs. 4b and 4c)
and the largest density enhancement occurs
at the two lateral dips of the new flux rope, close to its center 
(the left column of Fig. 5).

\subsection{Eruptive Case --- Inverse Prominence Configuration}

In the case of the IP configuration (Case 4b), the emerging
flux rope undergoes a two-stage evolution as we can see from Figure 2:
(i) a slow evolution (from 0 hrs to 7 hrs) sets in upon the flux rope emergence;
(ii) the flux rope then goes unstable at around 7 hrs (exhibiting a gradual 
acceleration) and later propagates upward into the interplanetary space 
with $\bar{v}=70.4$ km s$^{-1}$. The evolution of the magnetic field and 
plasma velocity, electric current density, and density enhancement for Case 
4b is shown in the right columns of Figures 3, 4, and 5, respectively. 
It is clearly noted that there is no reconnection 
on the leading edge of the flux rope because the magnetic field in the 
upper part of the rope runs in the same sense as the external field of 
the helmet dome. However, a vertical current sheet is formed,
trailing the flux rope from below as the rope rises (e.g., Figs. 4g and 4h).
Magnetic reconnection ensues and dissipates the current sheet.
As a result of reconnection as well as the rising X-type neutral point 
due to the fixed boundary condition at the inner boundary \citep{wu95},
the opened magnetic field is later reclosed low in the corona, 
as shown in Figures 3g and 3h. 
This enables the helmet streamer structure to recover at $\sim$ 40 hrs 
(not shown). Similar processes take place in Case 4a as well after a
vertical current sheet is formed below the old flux rope (Figs. 3c and 3d).
The recovery to equilibrium is much slower in the IP than in the NP
configuration. This is partly due to the different CME propagation speeds 
and dynamic time scales for these two magnetic topologies.
It is interesting to note that, in the early phase, 
the maxima of the electric current (Figs. 4e and 4f) 
and density enhancement (Figs. 5e and 5f) occur in 
the vertical current sheet and the lower part
of the emerging flux rope, respectively, in contrast with those of Case 4a.


\section{Discussions}

\subsection{Operating Forces}

In a MHD representation of the corona plasma, there are three forces, 
viz., the Lorentz force, pressure gradient force, 
and gravitational force, which fundamentally determine the dynamics 
of the flux rope system
(e.g., Wu et al. 1995; Guo et al. 1996; Hu \& Liu 2000). 
We show in Figure 6 the spatial distributions in the equatorial plane
of the normalized radial components of these forces 
for Cases 4a (left) and 4b (right) during the early stage of the simulation.
By examining the forces, we notice the following interesting features.

For the NP configuration (Case 4a), the pressure force is 
the dominant positive force to destabilize the helmet streamer in the early 
stage ($t < 4$ hrs) (Figs. 6a and 6b), but it drops significantly afterwards. 
This is true because 
the emerging flux rope carries substantial mass as well as upward momentum 
into the helmet dome and a large pressure force is accordingly developed;
after the emergence ($t > 4$ hrs) there is no more momentum being added 
into the corona except for the inner boundary conditions on the photosphere 
to maintain the background solar wind. 
In contrast, the Lorentz force in the newly formed flux rope
is relatively small. This happens for the following reasons.
On the one hand, the emerging flux rope carries a current in a direction
{\it opposite} to that in the external helmet-streamer field and magnetic reconnection
rapidly dissipates the current near the leading edge of the 
old flux rope around the equatorial plane.
As a result, the current density in that region is very small and 
the current distribution peaks on the flanks of the old flux rope 
(Figs. 4a, 4b, and 7a through 7c). 
On the other hand, because reconnection gradually removes the 
constraints of the helmet-dome's closed field 
ahead of the new flux rope, the new flux rope's material
and frozen-in magnetic field can rise and expand readily (Figs. 3a and 3b),
leading to a decreased magnetic field strength (Figs. 8a through 8c)
and further lowering the current density as well as the 
pressure gradient force inside the volume of the new flux rope.
Being the cross product of the current density and 
magnetic field strength, a small Lorentz force (Figs. 6a through 6c)
is therefore produced in the new flux rope and its vicinity, 
in the presence of the decreased current and magnetic field.

For the IP configuration (Case 4b), the pressure gradient force 
and gravitational force
exhibit similar behaviors and magnitudes as in Case 4a (Fig. 6). However, the 
Lorentz force tells us a different story. First, unlike Case 4a, 
the emerging flux rope in this case bears a current in the 
{\it same} direction as that of the helmet streamer above and this current 
produces an additional positive Lorentz force which can be understood in terms of
the attractive force between the two like-signed currents. Second, 
there is no reconnection on the leading edge of the flux rope (cf. Case 4a)
and the constraints from the overlying helmet-dome arcades cannot be readily 
removed. Therefore, the rise and expansion of the flux rope are suppressed
by the arcades. This leads to a pileup of plasma (e.g., Fig. 5e) in the flux rope 
during the course of the emergence and results in an enhanced magnetic field
(Figs. 8d through 8f) and current (Figs. 4e, and 7d through 7f). 
For these two reasons the Lorentz force appears much larger in this
case compared with that in Case 4a (Figs. 6d through 6f).
We also note that the Lorentz force flips its direction across the 
flux rope center due to the circulating
magnetic field: this force is negative (downward) in the upper half 
of the rope, tending to hold the rising material; it switches upward 
in the lower half, supporting the dense plasma (Figs. 6e and 6f).
This feature is less evident in Case 4a, as the Lorentz 
force is smaller and the topology is more complex 
(involving two flux ropes rather than one) in that case.
As to the temporal evolution, the flux rope experiences two stages
as we note in \S\ 3.2. Early during the emergence before the center of the flux rope
comes across the photosphere,
the pressure gradient force is the most dominant to lift off the rope plasma;
however, the Lorentz force remains negative above the center
and tends to suppress the upward motion (Fig. 6e). Later on, after the center
appears in the corona, the positive Lorentz force below it 
joins the pressure
force to work against the gravity as well as the confining Lorentz force
in the upper half of the rope. As the pressure force dies away after
the emergence completion, the positive Lorentz force takes over the
dominance. The complex interplay of these forces accounts for the
slow evolution stage ($<7$ hrs) of the flux rope which appears as a
hump in the speed-time profile (Fig. 2b). After that, the positive
Lorentz force in the lower part of the flux rope remains dominant,
responsible for the gradual acceleration of the flux rope starting 
at $t \sim 7$ hrs, and leads to its eventual eruption.

\subsection{Magnetic Topologies and Reconnection}

Now comes the question why the flux rope erupts faster in the NP 
than in the IP configuration.
We illustrate by simulation that in the NP 
configuration case magnetic 
reconnection plays a direct role in launching the CME (LZ02) 
by forming the new flux rope;
this process removes the closed field 
lines of the helmet dome as well as their constraints of the 
downward Lorentz force ahead of the new flux rope, 
thereby allowing the new flux rope to escape readily. 
Whereas, in the case of the IP configuration the 
overlying magnetic arcades tend to confine the flux rope; lacking mechanisms 
(e.g., reconnection) to remove the confinement, the flux rope thus fails to
reach a higher initial speed. 
To appreciate this point, let us further examine our simulation results,
focusing on the topological differences.

Firstly, the reader is reminded of
the physical environment specified by this numerical study. (i) The 
characteristic plasma $\beta$ at the bottom of the corona is unity, 
which means that the magnetic energy
and thermal energy are comparable, and the Lorentz force and pressure
gradient force are roughly of equal importance. (ii) A CME is initiated
by introducing a flux rope to emerge through the photosphere into the corona.
(iii) Except the magnetic polarity of the emerging flux rope, all the conditions
are the same for the two cases arranged in a pair 
(NP and IP configuration), which
implies that topology-independent forces (e.g., the pressure gradient and
gravitational force) would behave similarly in the two cases but 
topology-dependent forces (e.g., the Lorentz force) would not.

With these points in mind we realize that,
since the flux rope emergence injects a substantial amount of mass, 
magnetic flux, electric current and upward momentum into the corona, 
such a flux rope possesses a potential to disrupt the helmet streamer,
where the increasing pressure gradient force would play an important role
especially in the early stage of the emergence ($< 2$ hrs), 
such as in Cases 4a and 4b (Fig. 6). In the meantime, the inputted
mass also induces an extra downward gravitational force that competes with the
the upward pressure gradient force. 
The consequent evolution heavily depends on how the corona responds to
such an injection during the first a few hours. 
Since the pressure force and gravitational force are similar in
a pair of cases, we shall pay more attention on their different Lorentz force.

In Case 4a (NP configuration), for example, the flux rope emergence drives
magnetic reconnection on the leading edge of the flux rope. In turn,
reconnection clears the constraints ahead, i.e., the closed field lines
of the helmet dome, and thus allows the flux rope to rise more readily. This is 
a break-out situation similar to that of \citet{ant99}.
Once reconnection removes all the closed field lines,
the flux rope is left with an open-field channel ahead in the helmet streamer.
Note that the confining Lorentz force almost vanishes in the equatorial plane
above the center of the new flux rope (see Figs. 6a through 6c).  Therefore, 
the only remaining major constraints that would prevent the flux rope from erupting 
now come from the gravity. Driven by the large pressure gradient force, 
the flux rope can thus overcome the gravitational pull and readily escape 
along the open-field channel with a high initial speed.

Nevertheless, in the IP configuration case (e.g., Case 4b), 
there is no reconnection on the leading edge of the flux rope 
to remove the overlying arcades and the corona responds to the flux rope emergence 
with a negative Lorentz force above the flux rope center, much larger than
its counterpart in the corresponding NP configuration case 
(e.g., Figs. 6d through 6f). This confining Lorentz force, 
together with the gravity, overcomes part of the upward momentum 
and tends to suppress the emergence against the pressure gradient force.
Incidentally, it is worth mentioning that although the magnitude of the negative 
Lorentz force is not large in comparison with the more dominant
gravity, when the pressure force and gravity are in competition at 
some critical point, even a small additional force 
would easily turn the lever of force competition to one end. 
In this sense, the negative Lorentz force plays such a crucial role. 
This results in the hump-shaped portion in the speed-time profiles 
for the IP configuration cases (see Fig. 2b). Losing a significant amount of
momentum in the very early stage of the emergence, the flux rope
fails to be driven to a high initial speed even with the positive Lorentz
force coming into play later on.

The other NP and IP configuration cases show similar behaviors 
of the forces as in Cases 4a and 4b, respectively. As the radius
and emergence speed of the flux rope become smaller, less upward 
momentum is injected into the corona, and thus it is
more difficult to work against the gravitational pull. Depending
on the energetics of the emergence, the flux rope may erupt at a lower
speed (e.g., Cases 1a and 3b) or even fail to escape 
(e.g., Cases 1b and 2b).

In brief, in the NP configuration cases, the fast CMEs result from
the flux rope eruptions mainly driven by the pressure gradient force, with little
magnetic confinement; in the IP configuration cases, the slow CMEs are driven
by the pressure gradient force and positive Lorentz force in the lower
half of the flux rope, subject to the significant drags from
the confining Lorentz force above the flux rope center. This explains
the distinct speed-time profiles of CMEs in the two topologically different
types of cases.


\section{Concluding Remarks}

We have presented MHD simulations in the form of the flux rope emergence
to investigate the relationship 
between magnetic topologies and two-class CMEs as suggested 
by LZ02. In conclusion, our numerical results demonstrate that: 
(i) A CME-like expulsion is more readily produced in an NP 
than in an IP environment, under otherwise the same conditions.
This agrees with the results from the analytical calculations of magnetic energy 
storage in the two types of prominences given by \citet{zhanglow03}.
(ii) Early in the event, a CME originating from an NP environment 
tends to have a higher initial speed low in the corona while 
a CME from an IP environment tends to experience a slow evolution and then erupt 
with a lower initial speed and gradual acceleration.
(iii) One of the ratios of the average CME speeds for these two types of 
magnetic topologies is about 2.3, consistent with the observations 
reported by \citet{gos76} and \citet{cyr99}.
(iv) In an NP environment, magnetic reconnection occurs 
on the leading edge of the emerging flux rope.
This reconnection removes the magnetic confining force produced by
the closed external field ahead of the flux rope and launches a fast CME 
in a manner similar to the Magnetic Break-out Model \citep{ant99}. 
However, in an IP environment, with reconnection absent on the leading edge 
and subject to the magnetic confinement from the overlying arcades,
the emerging flux rope either fails to erupt (e.g., Cases 1b and 2b) 
or results in a slow CME (e.g., Case 4b), similar to the early works 
given by \citet{wu95} and \citet{wu97}.
Reconnection, taking place in the vertical current sheet
trailing the rising flux rope,
is a passive effect of the CME expulsion\footnote{Note that reconnection here 
does remove a small portion of the flux from the 
overlying, closed field in the helmet dome by converting it to the flux
in the outer layers of the flux rope.}.
In this sense, magnetic reconnection plays a principal role in generating 
a fast CME in the NP configuration but does not in the
IP environment \citep{zha02}.

On the whole, the present study qualitatively agrees well with the  
LZ02 scenario. However, for an NP configuration, 
the slingshot topology predicted by LZ02 is not present in our simulations
and the new flux rope formed by reconnection, as clearly shown in our results,
was not considered in their theory.
The reason to account for this difference may be explained as follows. 
To produce a slingshot topology (see Fig. 2 in LZ02), one must
expect that reconnection occurs at a single X-type neutral point
on top of the flux rope and as a result the pre-event, closed external
field lines of the helmet dome reconnect with the internal field lines of the
flux rope to form a ``slingshot". In our time-dependent simulations, 
as soon as the emerging flux rope touches the overlying helmet-dome 
field, a current sheet forms in between and the tearing mode instability may 
set in due to the prescribed finite resistivity. 
This may produce many small magnetic islands which may rapidly coalesce 
to form larger islands and eventually the new flux rope.
This new flux rope sits between the emerging rope and the helmet dome,
and on the flanks are the two X-type neutral points where subsequent
reconnection occurs. We propose that, depending on the reconnection development
which is difficult to predict in advance, both the slingshot topology 
(as in LZ02) and the topology with a new flux rope formed (as in our simulations) 
might be the case in a realistic solar environment, and both topologies
would be in favor of producing fast CMEs.

It should be pointed out that, although the CMEs are generated by emerging
flux ropes into the corona somewhat artificially in this numerical experiment, 
this paper is not aimed to investigate the initiation mechanisms of CMEs
which have been addressed elsewhere in the literature 
(e.g., Forbes 2000; Wu et al. 2000; 	
and references therein). However, 
regardless of CME initiations, the topology-dependent behavior
of the CME expulsion indeed reveals the logical connection between the
magnetic topologies and the two types of CME speed-height profiles, which
is the main goal of this study.

The CME speeds in our simulations are systematically lower than
observed values.  One of the reasons is that we traced the center
of the erupting flux rope to obtain the corresponding CME speed, which is
an underestimate because an observed CME speed is usually 
measured at the bright CME front that gains an additional speed due 
to the self-expansion of the flux rope relative to its center.
Another reason is that the initial energy content in the 
emerging flux rope is not large enough to launch a fast CME up to 800 km s$^{-1}$.
This deficiency could be remedied by seeking low-$\beta$ MHD 
solutions or by upgrading this 2-D model to a 2.5-D or a full 3-D one,
consequently increasing the energy content. These attempts would be beyond the 
scope of the present paper.
It is worth noting that \citet{wu04}, alternatively, have adopted 
a 2.5-D model and simulated both fast and slow CMEs (at speeds comparable to 
observed values) in an IP configuration, by invoking one or a combination of 
the three driving mechanisms: magnetic flux injection, mass drainage, 
and additional heating.  This suggests that the
two-class CME speed-height profiles could result from a variety of mechanisms 
among which the scenario of LZ02 was an initiative example. We would expect
that even faster CMEs will be produced if an NP configuration is considered by
\citet{wu04}.

Observations, as always, will be needed to assess a theory or a numerical study.
Several aspects in the LZ02 theory and in our simulations are observationally 
testable.  One prediction of the LZ02 theory is that magnetic reconnection, 
taking place either above or blow the flux rope,
plays distinct roles in the different field topologies of the two prominence environments. 
Solar flares, which are observational manifestations of reconnection, 
are hence expected to exhibit distinct timing and morphological 
behaviors in association with fast and slow CMEs.  Such behaviors have recently 
been reported by \citet{zha02} and \citet{zha03} using {\it TRACE} 
UV/EUV observations.  Hard X-ray (HXR) emissions produced by accelerated particles during flares
provide another tool to shed light on the magnetic topology of the reconnection 
site and the current sheet \citep{sui03}.  One can test whether there exist 
different morphologies of flares associated with the two types of CMEs, 
using HXR data obtained by {\it RHESSI} \citep{lin02} or 
the Hard X-ray Telescope onboard the {\it Yohkoh} satellite.
In particular, for limb flares, HXR images of the loop-top (LT) source, 
which is presumably located near/at the reconnection site,
can be compared with $H_{\alpha}$ images to check whether the LT
occurs above or below the erupting prominence (if any), and to see whether such 
an occurrence is associated with a fast or slow CME, respectively.

In the case of an NP environment,  a new flux rope is formed by reconnection
ahead of the old flux rope. If the remainder of the magnetic flux in the
old flux rope is significant, the CME would contain two flux ropes with
opposite chiralities of the magnetic field. This aspect of
our simulations may be tested by using {\it in situ} observations of the magnetic
configurations in the interplanetary counterparts of Earth-directed CMEs,
provided that the chiralities of the flux ropes are conserved during their 
interplanetary propagation. The January $10-11$, 1997 magnetic cloud (MC) 
observed by the {\it WIND} spacecraft contained
$^4$He$^{++}$/H$^{+}$ abundance similar to that of the streamer belt material,
suggesting an association between the MC and a helmet streamer.
In addition, a very cold region of exceptionally high density was detected at
the rear of the MC, and this dense region had an unusual
composition, indicating an association 
with the prominence material. This event was interpreted to be associated with
a CME on January 06, 1997 with an estimated speed of 450 km s$^{-1}$ 
observed by {\it SOHO}/LASCO \citep{bur98}. The very
cold region also contained a magnetic configuration, very likely of a flux rope, 
though its size was much less than the MC. We suggest that,
by respectively fitting the observed magnetic field of the MC and the very cold region 
with a flux rope model, their chiralities can be determined and thus will
provide observational evidence to test our simulation results.
Recently, multiple MCs have been reported by \citet{wan03}. 
The March $03-05$, 2001 MCs, one of the three events 
in their study, consisted of two flux ropes with opposite chiralities, 
one right-handed and the other left-handed.
They interpreted these two MCs as a consequence of two successive CMEs with projected
speeds of 313 km s$^{-1}$ and 631 km s$^{-1}$, respectively. 
However, identifying the solar origin of a MC has its uncertainty 
and it was possible that the two MCs might result from a single fast CME. 
If this is the case, such events would provide observational supports to our model.
It will be interesting to see, statistically, whether two successive MCs are
associated with a single CME or successive CMEs and we look forward to such 
kind of observational tests to check on our simulations.

Ultimately, observations of the magnetic field will provide more incisive
information to verify or reject the LZ02 theory. 
A study using line-of-sight MDI magnetograms is in progress in order to
determine if there are differences in the magnetic environments of
the source regions of the two types of CMEs 
(M. Zhang \& J. Burkepile, private communication).
In the meantime, advance in measuring the magnetic field in the corona 
by polarimetric methods \citep{lin98,jud98,lin00,tru01,tru02} 
will be able to put the theory to a direct test in the near future.

\begin{acknowledgements}

Work performed by WL, XPZ and PHS was 
supported at Stanford University by NASA grants NAGW 2502 and NAG5-10483, NSF 
grant ATM 9400298, and ONR grant N0014-97-1-0129. Work performed by STW 
was supported by NASA grant NAG5-12843 and NSF grant ATM-0316115.

\end{acknowledgements}



\clearpage

\begin{figure}
\centering
\includegraphics[width=14cm]{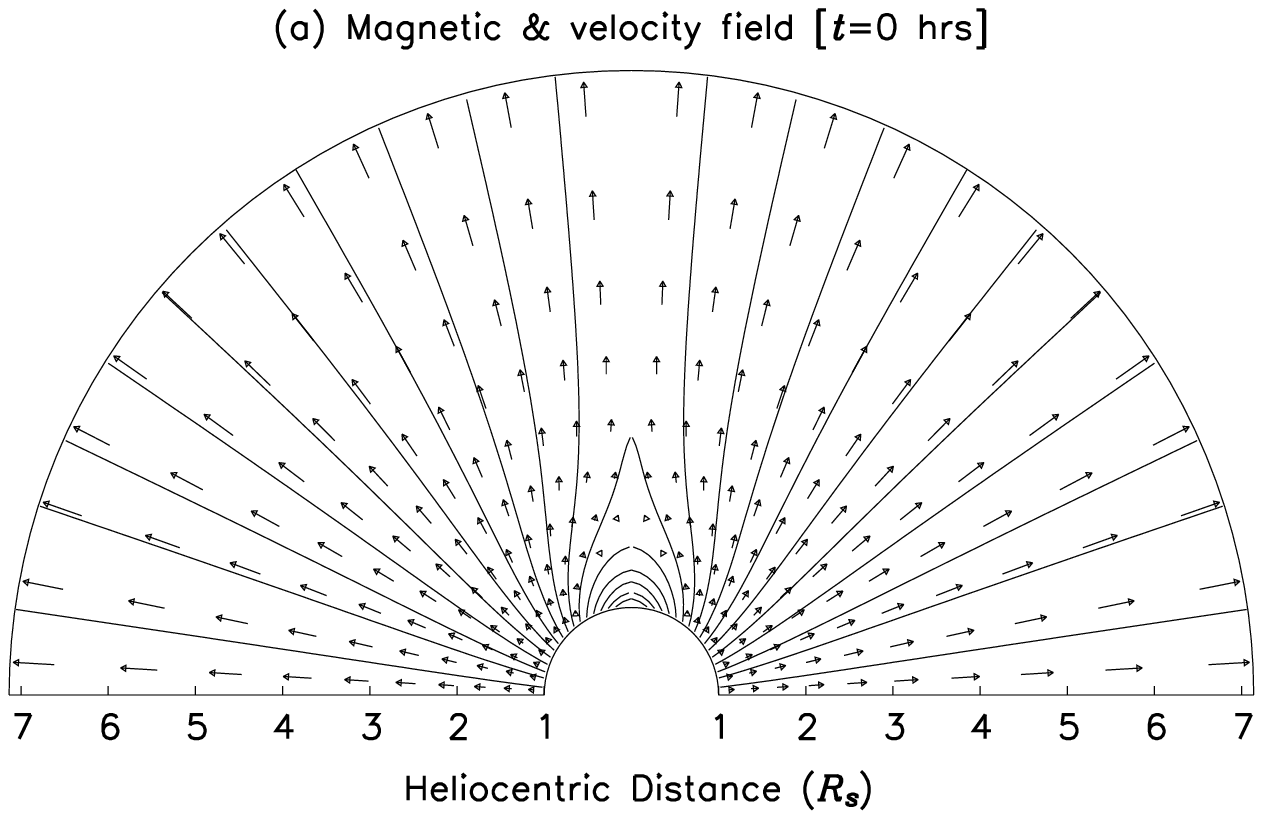}
\includegraphics[width=14cm]{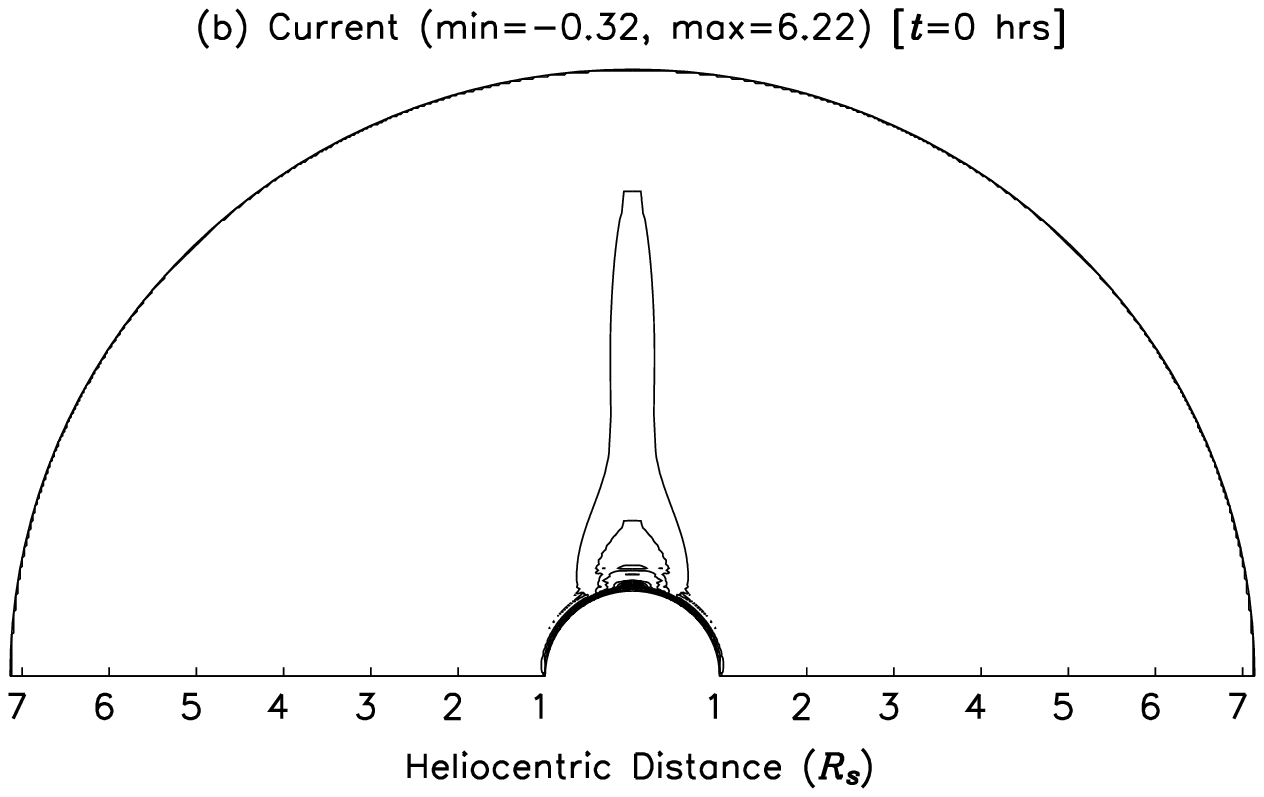}
\caption{The initial state of the solar corona with a helmet streamer over the equator: 
(a) the magnetic field lines and solar wind velocity; 
(b) spatial distribution of the azimuthal current density $J_{\phi}$ (in contours). 
The minimum and maximum are shown on top of Panel b in units of 
$J_0=2.29 \times 10^{-7} $ A m$^{-2}$ and 18 contour levels are uniformly
set between these extrema. Note the vertical current sheet extending upward from 
the helmet streamer cusp.
\label{fig1}}
\end{figure}


\begin{figure}
\centering
\includegraphics[width=14cm]{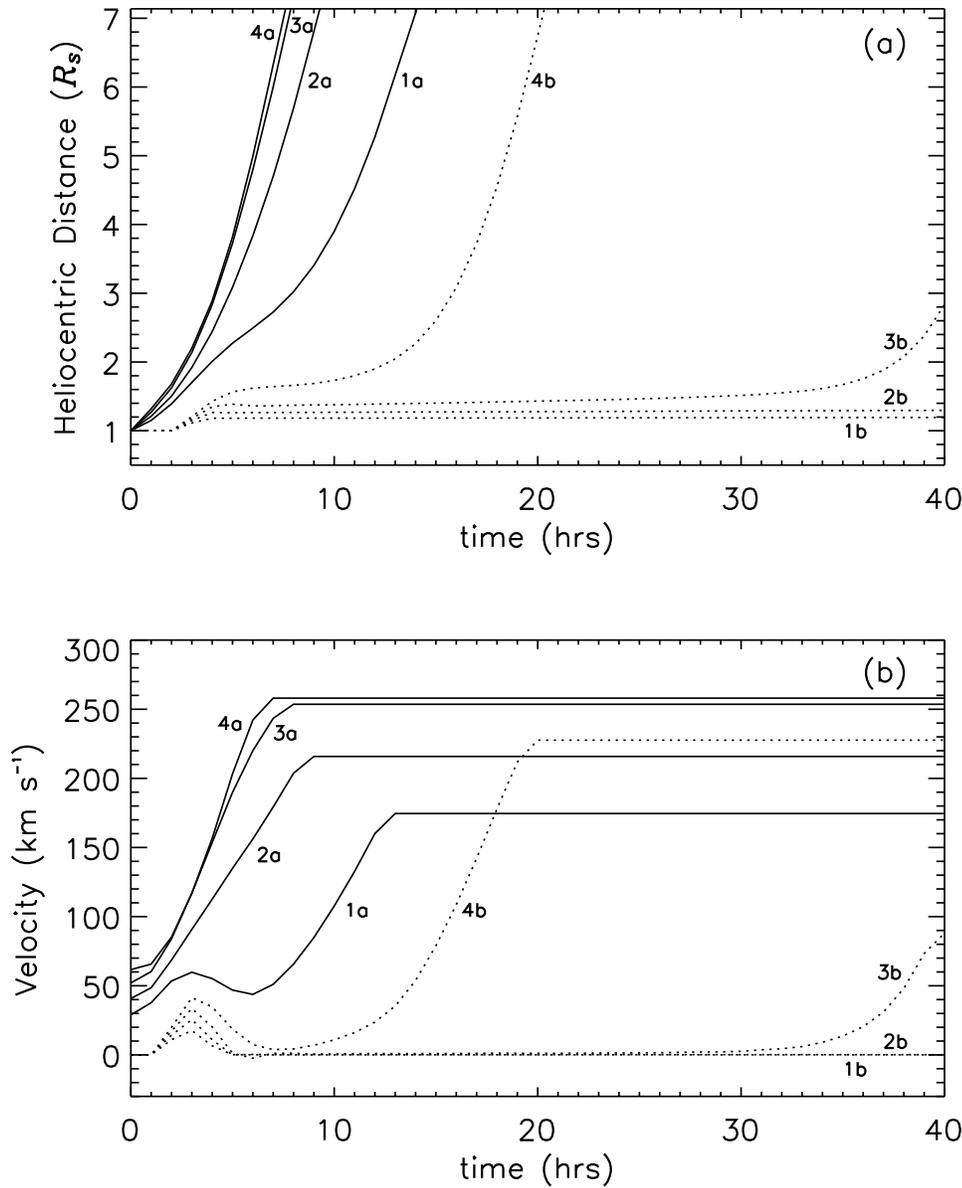}
\caption{(a) Height-time and (b) speed-time profiles 
of the center of the ``erupting flux rope"$^1$ for the eight 
cases listed in Table 2. The solid (dotted) lines correspond to the cases with the
NP (IP) configuration. Note that for the speed profiles 
in Panel b: the initial humps at $t \sim 3$ hrs are results of the flux rope emergence; 
the final flat portions (except for Cases 1b and 2b) are extrapolations of 
the speeds evaluated at the outer boundary ($r=7.14 R_s$); also the final 
portions of curves 1b and 2b overlap each other.
\label{fig2}}
\end{figure}


\begin{figure}
\centering
\includegraphics[width=17cm]{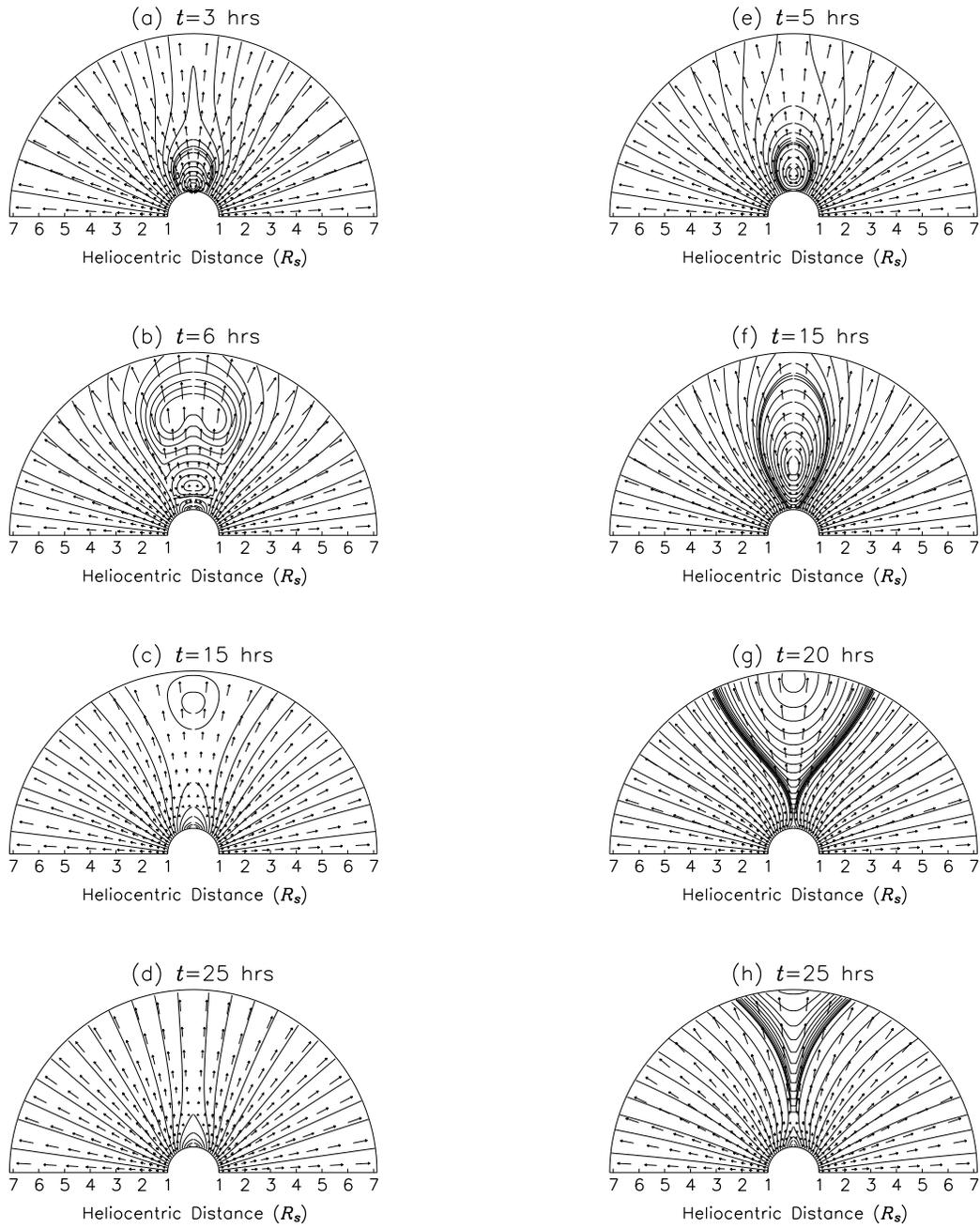}
\caption{Temporal evolution of the magnetic field and plasma flow velocity 
for Case 4a (NP configuration, left column) and 
Case 4b (IP configuration, right column).
\label{fig3}}
\end{figure}

\begin{figure}
\centering
\includegraphics[width=17cm]{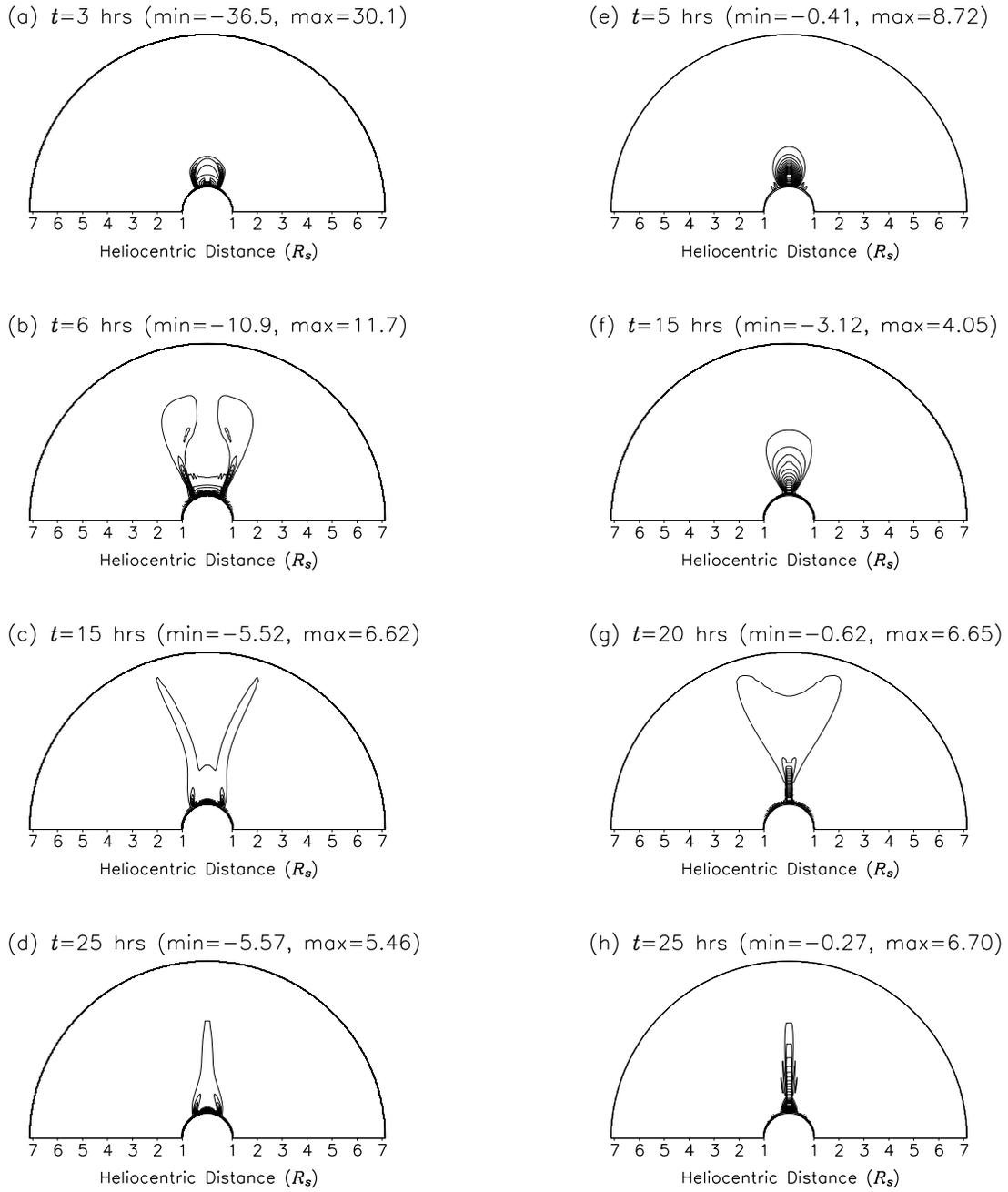}
\caption{The corresponding contours of the azimuthal electric current density $J_{\phi}$ 
(in units of $J_0=2.29 \times 10^{-7} $ A m$^{-2}$) of Figure 3.
The minima and maxima are shown on top of each panel and the intervals 
[min, max] are evenly divided into 38 and 34 contour levels for Cases 4a (left)
and 4b (right), respectively.
\label{fig4}}
\end{figure}

\begin{figure*}
\centering
\includegraphics[width=17cm]{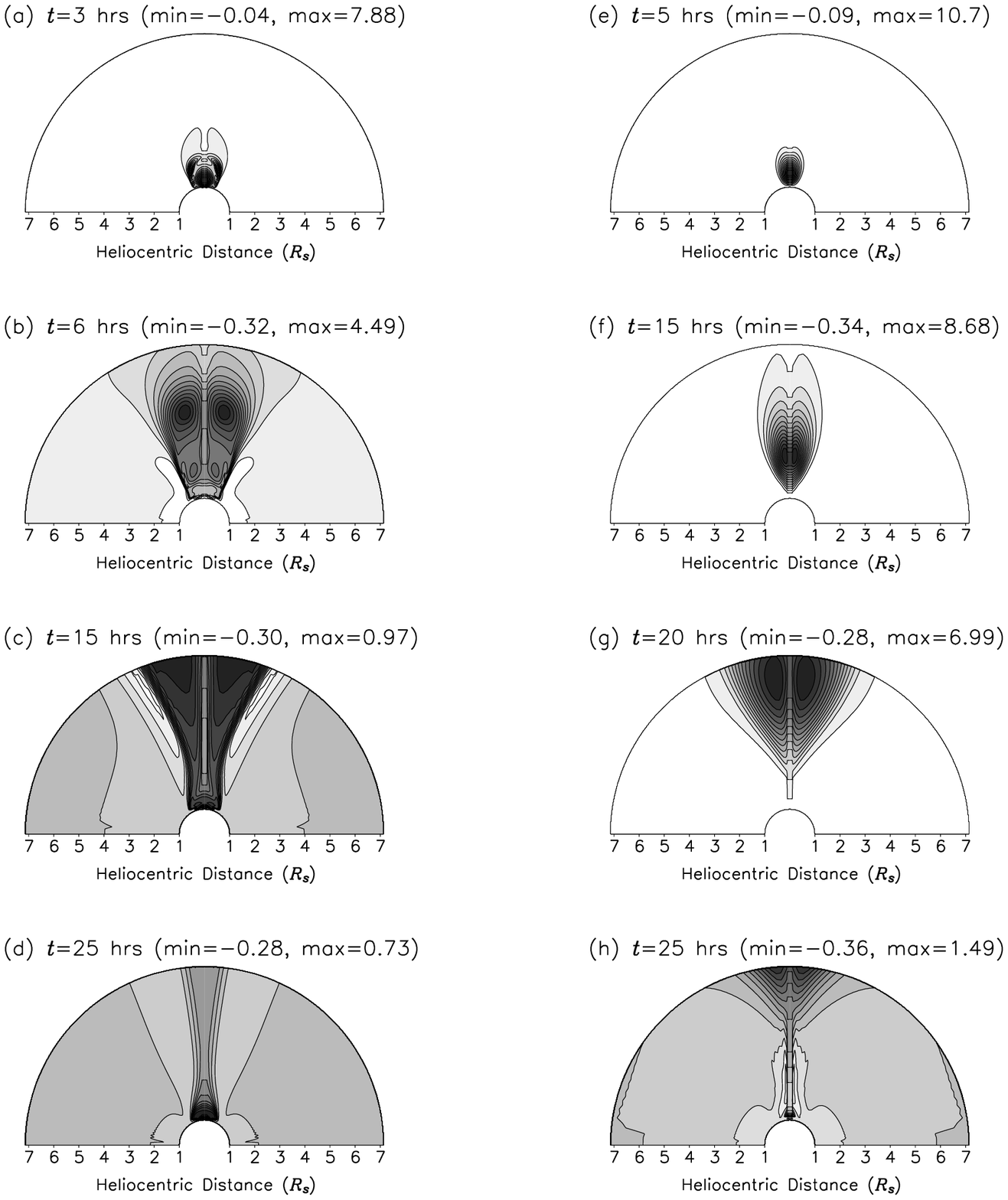}
\caption{The corresponding relative density enhancement, 
i.e., $[\rho(r,\theta,t)-\rho(r,\theta,0)] / \rho(r,\theta,0)$,
of Figure 3, plotted in the same manner as Figure 4 
except that here are 15 contour levels filled with gray scale shades
(bright: low, dark: high).  
\label{fig5} }
\end{figure*}

\begin{figure}
\centering
\includegraphics[width=16cm]{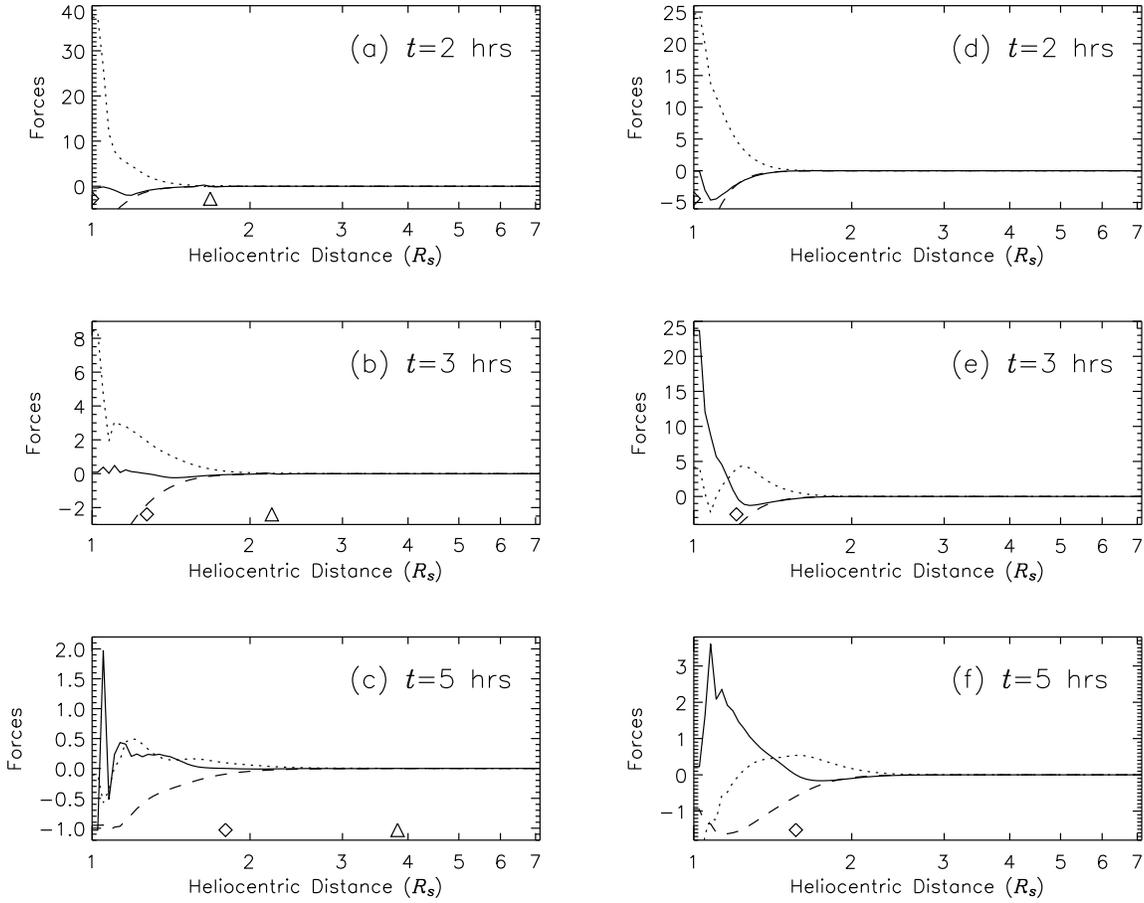}
\caption{Radial components of the forces 
(i.e., Lorentz force ------, pressure gradient force ......, 
and gravitational force - - -) per unit volume as functions of the 
heliocentric distance in the equatorial plane at various times for 
the NP (Case 4a, left column) and IP (Case 4b, right column) 
configuration. Distance is logarithmically plotted to show details 
near the solar surface. All the forces are normalized by the gravitational
force at the equator on the photosphere in the initial state. 
Note the flux rope center (i.e., the O-type neutral point) in each case.
Diamond symbols denote the radial positions of the 
center of the originally emerging flux rope;
triangles mark those of the new flux rope formed by reconnection 
in the case of the NP configuration.
\label{fig6}}
\end{figure}

\begin{figure}
\centering
\includegraphics[width=16cm]{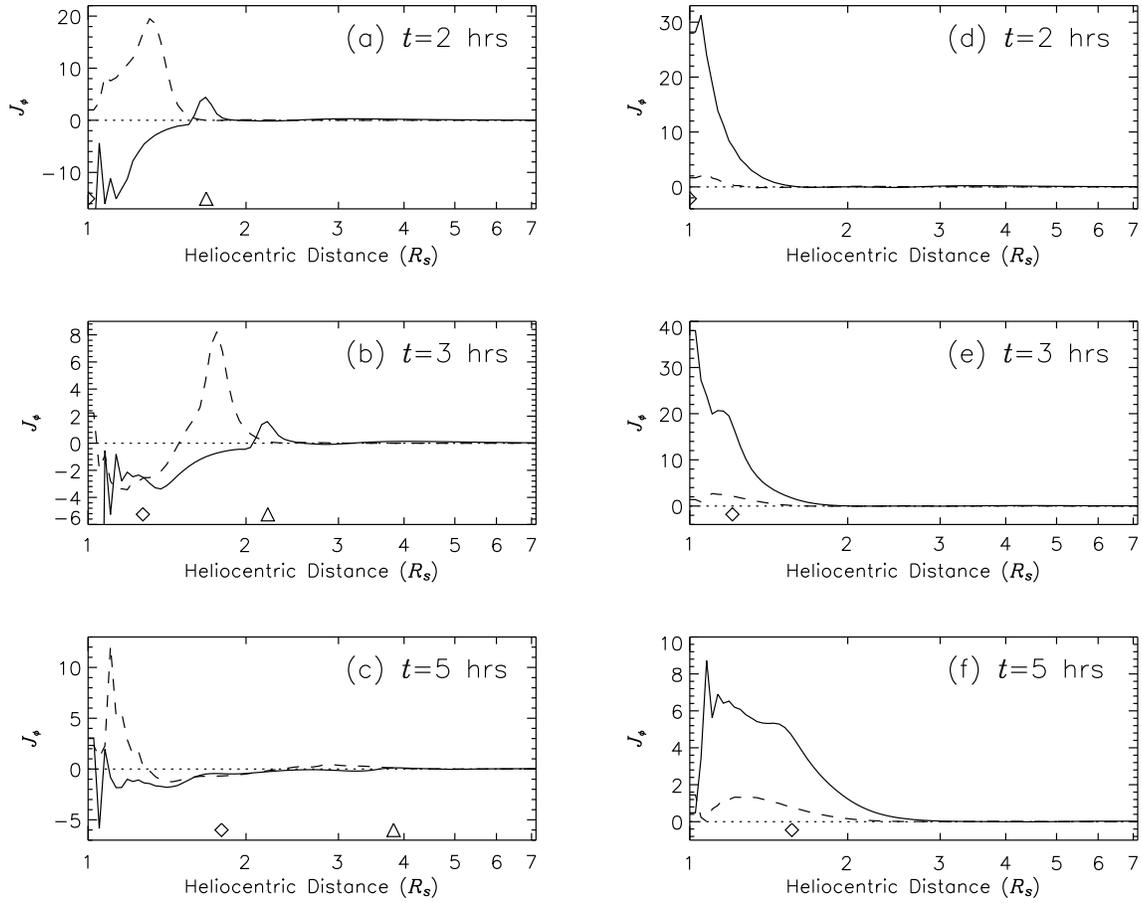}
\caption{Azimuthal current density distribution, $J_{\phi}$ 
(in units of $J_0=2.29 \times 10^{-7} $ A m$^{-2}$)
at various times corresponding to Figure 6, 
in the equatorial plane ($\theta=90^\circ$, solid lines) 
and along $\theta=72^\circ$ (or $\theta=108^\circ$, dashed lines) 
where the split current sheets in Case 4a are roughly located.
The dotted line in each panel marks the zero level.
\label{fig7}}
\end{figure}

\begin{figure*}
\centering
\includegraphics[width=16cm]{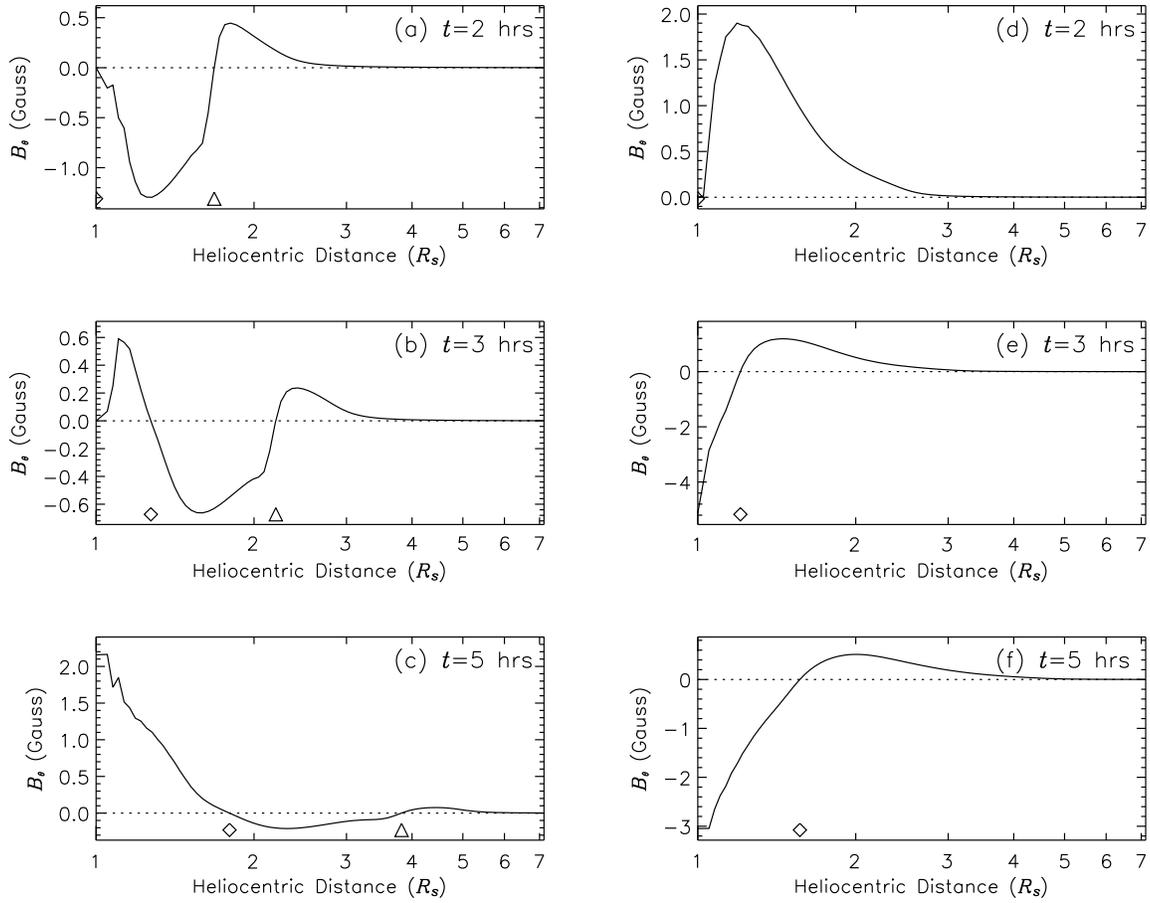}
\caption{$\theta$-component of the magnetic field ($B_\theta$) 
vs. heliocentric distance
in the equatorial plane ($\theta=90^\circ$) 
at various times, corresponding to Figure 6.
The dotted lines mark the zero levels. Note that $B_r$ vanishes at $\theta=90^\circ$
under the symmetric boundary condition.
\label{fig8}}
\end{figure*}


\clearpage

\begin{table}[ht]
\caption{Characteristic parameters of the initial state.}
\vspace {0.05in}
\begin{tabular}{ll}
\hline\hline
 $n_0$ (electron number density) & $3.2\times 10^8$ cm$^{-3}$\\
 $T_0$ (temperature) & $1.8\times10^6$ K\\
 $B_0$ (magnetic induction strength) & $2.0$ G\\
 $\beta_0$ (plasma $\beta$) & 1.0 \\
 $c_s$ (sound speed) & 176.7 km s$^{-1}$\\
 $v_A$ (Alfv\'{e}n speed) & 243.9 km s$^{-1}$\\
 $v_{sw}$ (solar wind speed) & 209.7 km s$^{-1}$\\
\hline
\end{tabular}

{Note. --- All the quantities refer to the condition in the 
equatorial plane. The solar wind speed is evaluated at the outer boundary
and the others at the inner boundary.}

\end{table}

\clearpage

\begin{table}[ht]
\caption{Characteristics of studied cases.}
\renewcommand{\footnoterule}{}
\vspace {0.05in}
\begin{tabular}{llcllll}
\hline\hline
Cases & $a^{(a)}$  & Field$^{(b)}$ & Energy$^{(c)}$ 
& $v_{em} ^{(d)}$  & $v_f^{(e)}$    & $\bar{v}^{(f)}$  \\
& ($R_s$) &  & ($ 10^{31}$ ergs) & (km s$^{-1}$) 
& (km s$^{-1}$)
&(km s$^{-1}$) \\

\hline
1a & 0.10 & NP & 1.76 & 9.7  & 174.6 & 90.0 \\
2a & 0.15 & NP & 3.95 & 14.5 & 215.8 & 133.9 \\
3a & 0.20 & NP & 7.02 & 19.3 & 253.6 & 152.6 \\
4a & 0.25 & NP & 11.0 & 24.2 & 258.0 & 161.0 \\
\hline

1b & 0.10 & IP & 1.76 & 9.7  & 0.1   & 1.0  \\
2b & 0.15 & IP & 3.95 & 14.5 & 0.2   & 1.5 \\
3b & 0.20 & IP & 7.02 & 19.3 & 88.0  & 9.3 \\
4b & 0.25 & IP & 11.0 & 24.2 & 227.6 & 70.4 \\
\hline
\end{tabular}

\begin{list}{}{}
\item[$^{(a)}$]{The radius of the emerging flux rope.}
\item[$^{(b)}$]{Magnetic field topologies in terms of NP or IP configurations.}
\item[$^{(c)}$]{The combination of the thermal and magnetic energy in the volume 
 of the emerging flux rope, assuming the third-dimensional thickness 
 $\Delta z' = 0.1 R_s$.}
\item[$^{(d)}$]{The flux rope emergence speed in Step 2 as described in \S\ 2.}
\item[$^{(e)}$]{The final speed of the center of the ``erupting flux rope"$^1$.
 This speed is evaluated at the end of the simulation (40 hrs) 
 or at the time when the flux rope center reaches the outer boundary 
 of the computational domain, whichever occurs first.}
\item[$^{(f)}$]{The average speed of the ``erupting flux rope"$^1$ center
  over the time interval during which the center remains in the computational domain.}
\end{list}

\end{table}

\end{document}